\RequirePackage{fix-cm}
\documentclass[twocolumn]{svjour3}          
\smartqed  
\usepackage{graphicx}
\usepackage{amsmath, amssymb, amsfonts, nccmath} 
\usepackage{braket}
\usepackage{graphicx}
\usepackage{caption}
\usepackage{cancel}
\usepackage{pifont}
\usepackage{mathtools}
\usepackage{bm}
\usepackage{empheq}
\usepackage{subfig}
\usepackage{fancyhdr}
\usepackage{setspace}
\usepackage{geometry}
\usepackage{cases}
\usepackage{xcolor}
\def\bbone{\scalebox{0.8}{$\mathbb{I}$}}

\newcommand*\rfrac[2]{{}^{#1}\!/_{\!#2}}
\definecolor{light-gray}{gray}{0.9}

\usepackage{tikz}

\begin{document}

\title{From a quantum theory to a classical one
}

\author{A. Coppo         \and
        A. Cuccoli \and
	C. Foti \and
	P. Verrucchi
}
\institute
{A. Coppo \at 
Dipartimento di Fisica ed Astronomia Universit\`a di Firenze and Istituto Nazionale di Fisica Nucleare,
via G. Sansone 1, I-50019 Sesto Fiorentino (Italy) \\
           \and
A. Cuccoli \at
Dipartimento di Fisica ed Astronomia Universit\`a di Firenze and Istituto Nazionale di Fisica Nucleare,
via G. Sansone 1, I-50019 Sesto Fiorentino (Italy) \\
\and
C. Foti \at
Dipartimento di Fisica ed Astronomia Universit\`a di Firenze and Istituto Nazionale di Fisica Nucleare, 
via G. Sansone 1, I-50019 Sesto Fiorentino (Italy) \\
\and
P.Verrucchi \at
Istituto dei Sistemi Complessi-CNR,
Dipartimento di Fisica ed Astronomia Universit\`a di Firenze, and Istituto Nazionale di Fisica Nucleare,
via G. Sansone 1, I-50019 Sesto Fiorentino (Italy) 
}

\date{Received: date / Accepted: date}
\maketitle
\begin{abstract}

We present and discuss a formal approach for describing the quantum to classical crossover based on the group-theoretic construction of generalized coherent states. 
The method was originally introduced by L. Yaffe\cite{Yaffe82} in 1982 for tackling large-$N$ quantum field theories, and has been recently used for studying open 
quantum systems whose environment, while becoming macroscopic, may or may not display a classical behaviour\cite{2015LCV,2017RFCTVP,2019FHMV,Coppo19MThesis}. Referring to these recent developments, in this paper we
provide the essential elements of Yaffes's approach in the framework of standard quantum mechanics, so as to clarify how the approach can be used without referring to quantum field theory.
Moreover, we address the role played by a possible global symmetry in making the large-$N$ limit of the original quantum theory to flow into a formally well defined classical theory, and we specifically consider the quantum-to-classical crossover of angular momentum. We 
also give details of a paradigmatic example, namely that of $N$ free one-dimensional spinless particles. Finally, we discuss upon the foundational requirement that 
any classical description should ultimately be derived from an underlying quantum theory, that however is not, and should never be confused with, the one obtained 
via some quantization procedure of the classical description itself.

\end{abstract}

\section{Introduction}
\label{intro}
Progresses in quantum technologies have recently made necessary to deeply understand the relation between macroscopic
objects that behave according to a classical theory, and the quantum world of microscopic systems, in order to find the best strategies for using, interacting, and 
exerting control upon small and fragile quantum devices. Key to this understanding is a formal description of the so called quantum to classical crossover, 
implying the possibility of connecting the geometrical structure of classical physics with the algebraic one featured by quantum mechanics. Some powerful tools in 
this framework can be found in the literature relative to the so called large-$N$ Quantum Field 
Theories: although they cannot be straightforwardly used in different settings, such as those typically arising in 
the analysis of open quantum systems, where the system undergoing the above crossover is just the big partner 
of a small quantum object, they are versatile enough to be adapted and turn very useful even in these frameworks.
In particular, the way L. Yaffe\cite{Yaffe82} in 1982 tackled some large-$N$ quantum field theories, has demonstrated very powerful and has been recently used for 
studying open quantum systems whose environment, while becoming macroscopic, may or may not display a classical behaviour
\cite{2015LCV,2017RFCTVP,2019FHMV,Coppo19MThesis}.
In this paper, after providing the essential elements of Yaffes's approach in the framework of standard quantum mechanics, we elaborate upon the role of the global symmetry, whose presence in the original quantum theory turns out to be a primary requirement to ensure that its large-$N$ limit is a  well defined classical theory. The practical implementation of the general abstract approach is described in detail for two specific examples: the quantum-to-classical crossover of angular momentum, and the deduction of the classical limit of a system made of $N$ free one-dimensional spinless particles.
The structure of the paper is as follows:
In Sec.\ref{sec:GCS} we introduce the Generalized Coherent States (GCS, which are essential in Yaffe's procedure) 
via the group-theoretical approach, independently developed by Gilmore~\cite{Gilmore72} and 
Perelomov~\cite{Perelomov72} in 1972. Following Ref.~\cite{ZhangFG90} we describe the 
algebraic procedure to construct GCS starting from the knowledge of the dynamical group of the system. In particular, we show how to construct GCS for systems 
associated to one of the two real forms of the Lie group $SL(2,\mathbb{C})$, namely the non-compact one $SU(1,1)$, whose proper GCS are the so called
Pseudo-spin Coherent States (PCS).
In Sec.\ref{sec:From_quantum to_classic} we identify the conditions ensuring that a quantum theory has a well defined classical limit, while in Sec.\ref{O(N) vector model} we consider a specific case to show that such limit can be obtained by increasing the number of degrees of freedom $N$ of the original quantum theory, i.e. when the system it 
describes becomes macroscopic, as briefly discussed in the last concluding section.  
\section{Generalized Coherent States}
\label{sec:GCS}
Any quantum theory $\mathcal{Q}$ can be defined in terms of an algebra, possibly a Lie algebra $\mathfrak{g}$, and a Hilbert space $\mathcal{H}$, which is the
carrier space of an irreducible representation of $\mathfrak{g}$. All the physically relevant operators on $\mathcal{H}$, except for the propagators, are elements of
such representation. On the other hand, according to the evolution-postulate of quantum mechanics, the propagators of $\mathcal{Q}$ are elements of a unitary
irreducible representation of the Lie group $\mathcal{G}$ obtained from $\mathfrak{g}$ via a Lie exponential map, for that $\mathcal{G}$ is called ``dynamical
group''. In what follows, for the sake of a lighter presentation, we will most often identify algebras and groups with their respective representations. 

Let us now consider a  generic quantum system described by a theory $\mathcal{Q}$ such that its Hamiltonian $\hat{H}$ belongs to $\mathfrak{g}$:
\begin{equation}\label{generic hamiltonian}
\hat{H}{=}H(\hat{g}_i)~,~\mbox{with}~~\hat{g}_i \in{\mathfrak{g}}~,~\mbox{and}~~[\hat{g}_i,\hat{g}_j]= c_{ij}^k \hat{g}_k~.
\end{equation}
If we limit our analysis to semisimple Lie algebras (or any algebra admitting a Cartan decomposition), the Cartan basis \{$\hat{D_i},\hat{E}_\alpha,\hat{E}_{-\alpha}$\} is defined, with
\begin{eqnarray*}
& &[\hat{D}_i,\hat{D}_j]=0~,~[\hat{D}_i,\hat{E}_\alpha]=\alpha_i \hat{E}_\alpha~,\\
& &[\hat{E}_\alpha,\hat{E}_{-\alpha}]=\alpha_i \hat{D}^i~,~[\hat{E}_\alpha,\hat{E}_\beta]=C_{\alpha\beta}\hat{E}_{\alpha+\beta}~,
\end{eqnarray*}
$\hat{D}_i$ Hermitian ($\hat{D}_i^\dagger=\hat{D}_i$), and $\hat{E}_\alpha$ such that $\hat{E}_\alpha^\dagger=\hat{E}_{-\alpha}$.The elements $\hat{D}_i(\hat{E}_\alpha)$ are dubbed {\it diagonal}({\it shift}) operators.
\\
Once a normalized reference state $\Ket{\Phi_0}$ in $\mathcal{H}$ is chosen, usually so as to be both an eigenstate of diagonal operators and a maximal weight state, 
i.e.
\begin{equation}\label{reference state hp}
\begin{cases}
\hat{D}_i\Ket{\Phi_0}=d_i\Ket{\Phi_0} \;\;\; d_i \in \mathbb{R}~,\\
\hat{E}_\alpha\Ket{\Phi_0}=0\;\;\;      \forall\alpha>0~,\\
\Braket{\Phi_0 | \Phi_0}=1~,
\end{cases}
\end{equation}
one can identify the subgroup $\mathcal{F}\subset \mathcal{G}$ that leaves $\Ket{\Phi_0}$ invariant up to a phase factor, 
i.e. $\hat{F}\in \mathcal{F} \rightarrow \hat{F}\Ket{\Phi_0}=\Ket{\Phi_0}e^{i\varphi(\hat{F})}$; this subgroup is called stabilizer of $\mathcal{G}$ 
with respect to $\Ket{\Phi_0}$. Finally, referring to the coset $\mathcal{G}/\mathcal{F}$, the generalized coherent states $\Ket{\Omega}$ are defined by:
\begin{equation}\label{displacement operator}
\hat{G}\Ket{\Phi_0}=\hat{\Omega}\hat{F}\Ket{\Phi_0}=\hat{\Omega}\Ket{\Phi_0}e^{i\varphi(\hat{F})}:=\Ket{\Omega}e^{i\varphi(\hat{F})}
\end{equation}
where
\begin{equation*}
\hat{G}=\hat{\Omega} \hat{F} \;\;\; \hat{G} \in \mathcal{G},\;\hat{F}\in \mathcal{F}, \; \hat{\Omega}\in \mathcal{G}/\mathcal{F}~.
\end{equation*}
We notice that GCS are in one-to-one correspondence with the elements $\hat{\Omega}$ of $\mathcal{G}/\mathcal{F}$.

\subsection{Differential structure of $\mathcal{G}/\mathcal{F}$}
\label{ssec:Diff_structure}
According to the ``quotient manifold theorem'' \cite{Lee12},
the coset $\mathcal{G}/\mathcal{F}$ can be associated to a complex manifold  $\mathcal{M}$ whose points $\Omega$ are in one-to-one correspondence with operators
$\hat{\Omega}$ in $\mathcal{G}/\mathcal{F}$, and hence with the states $\Ket{\Omega}$. Since the algebra $\mathfrak{g}$ is semisimple, it satisfies the Cartan
decomposition in the form $\mathfrak{g}=\mathfrak{f}\oplus\mathfrak{p}$, where $\mathfrak{f}$ is the algebra of $\mathcal{F}$ and $\mathfrak{p} = \xi^\beta
\hat{E}_\beta-{\xi^\beta}^* \hat{E}_{-\beta}$ is its orthogonal complement; therefore we can use the coordinates $\left\{\xi^\beta, {\xi^\beta}^*\right\}$ to
write
\begin{equation}\label{generalized displacement operator}
\hat{\Omega}=\mathrm{e}^{\left(\xi^\beta \hat{E}_\beta-{\xi^\beta}^* \hat{E}_{-\beta}\right)},\qquad \xi_{\beta}\in\mathbb{C}~.
\end{equation}
One can use other coordinate-systems such as, dropping the $\beta$-index for the sake of a lighter notation,
\begin{equation}\label{zeta coordinates}
\begin{dcases}
\zeta=\xi\frac{\sin{\sqrt{\xi^\dagger\xi}}}{\sqrt{\xi^\dagger\xi}}\;\; \mbox{if $\mathcal{M}$ is compact}~,\\
\zeta=\xi\frac{\sinh{\sqrt{\xi^\dagger\xi}}}{\sqrt{\xi^\dagger\xi}}\;\; \mbox{if $\mathcal{M}$ is not compact}~,
\end{dcases}
\end{equation}
or the one yielding a complex projective representation,
\begin{equation}\label{tau coordinates}
\begin{cases}
\tau=\zeta(1-\zeta^\dagger \zeta)^{-\frac{1}{2}}\;\; \mbox{if $\mathcal{M}$ is compact}~,\\
\tau=\zeta(1+\zeta^\dagger \zeta)^{-\frac{1}{2}}\;\; \mbox{if $\mathcal{M}$ is not compact}\,.
\end{cases}
\end{equation}
\subsection{Metric and measure}
\label{ssec:Metric_and_measure}
It can be demonstrated \cite{Hua63} that $\mathcal{M}$ is endowed with a natural metric that can be expressed in the $\tau$ coordinates as
\begin{equation}\label{natural metric}
\begin{aligned}
&ds^2=g_{\alpha\beta}d\tau^\alpha d{\tau^\beta}^*\;\; \mbox{where} \;\; g_{\alpha\beta}:=\frac{\partial^2F(\tau,\tau^*)}{\partial\tau^\alpha \partial{\tau^\beta}^*}~,\\
&F(\tau,\tau^*)=\log{N(\tau,\tau^*)},\;\; N(\tau,\tau^*)=\Braket{\tilde{\tau} | \tilde{\tau}}~,\\ 
&\Ket{\tilde{\tau}}=e^{\tau^\beta \hat{E}_\beta}\Ket{\Phi_0}~,
\end{aligned}
\end{equation}
where $\Ket{\tilde{\tau}}$ is a non-normalized GCS.
This allows one to get information upon the manifold $\mathcal{M}$.
Through $ds^2$ one can define a canonical volume form on $\mathcal{M}$,  i.e. a measure
\begin{equation}\label{natural measure}
d\mu(\Omega)= \mbox{const} \times \det(g) \prod_{\alpha} d\tau^\alpha d{\tau^\alpha}^*~.
\end{equation}

\subsection{Overcompleteness of coherent states}
\label{ssec:Overcompleteness_of_CS}
Using $d\mu(\Omega)$, GCS are demonstrated to form an overcomplete set of states on $\mathcal{H}$, providing a continuous resolution of the identity,
i.e.
\begin{equation}\label{coherent state completeness}
\hat{\mathbb{I}}=\int_{\mathcal{G}/\mathcal{F}} {d\mu(\Omega)\,\Ket{\Omega}\Bra{\Omega}}~.
\end{equation}
The prefix ``over'' in the adjective 'overcomplete' indicates that coherent states are ``a lot'': in fact, despite being normalized,
$\Braket{\Omega | \Omega}=\Bra{\Phi_0}\hat{G}^{-1}\hat{G}\Ket{\Phi_0}= \Braket{\Phi_0 | \Phi_0}=1$, $\forall\hat{G}\in \mathcal{G}$, they are not orthogonal,
\begin{equation}
\begin{aligned}
&\Braket{\Omega | \Omega'}=\Bra{\Phi_0}\hat{\Omega}^{-1}\hat{\Omega}'\Ket{\Phi_0}=\\
&=\Bra{\Phi_0}\hat{G}^{-1}\hat{G}'\Ket{\Phi_0}\mathrm{e}^{i\varphi}=\Bra{\Phi_0}\hat{G}''\Ket{\Phi_0}\mathrm{e}^{i\varphi}\neq 0~,
\end{aligned}
\end{equation}
$\forall\;\hat{G},\hat{G}',\hat{G}'' \in \mathcal{G}$, and $\hat{\Omega},\hat{\Omega}'\in \mathcal{G}/\mathcal{F}$.

\subsection{Symplectic structure}
\label{ssec:Symplectic_structure}
$\mathcal{M}$ is equipped with a symplectic structure that allows one to identify it as a phase space, possibly the one proper to the classical system into which the 
original quantum system flows when the classical limit is rigorously performed. The symplectic form on $\mathcal{M}$ has the coordinate representation
\begin{equation}\label{symplectic form}
\omega=-i\;\sum_{\alpha\beta}g_{\alpha\beta}\,d\tau^\alpha\wedge d{\tau^\beta}^*~,
\end{equation}
and it is used to define the Poisson brackets
\begin{equation} 
\left\{f,g\right\}_{PB}:=i\;\sum_{\alpha\beta}g^{\alpha\beta}\left(\frac{\partial f}{\partial \tau^\alpha} \frac{\partial g}{\partial {\tau^\beta}^*} - 
\frac{\partial g}{\partial \tau^\alpha} \frac{\partial f}{\partial {\tau^\beta}^*}\right).
\end{equation}
Switching to the $\zeta$ coordinates, and defining $w$ and $v$ via
\begin{equation}\label{w,v coordinates}
\zeta_\beta=\frac{1}{\sqrt{2}}\left(w_\beta-iv_\beta\right),\quad \zeta_\beta^*=\frac{1}{\sqrt{2}}\left(w_\beta+iv_\beta\right)~,
\end{equation}
one obtains the Poisson brackets in the standard form,
\begin{equation} 
\left\{f,g\right\}_{PB}=\sum_{\alpha}\left(\frac{\partial f}{\partial v^\alpha} \frac{\partial g}{\partial w^\alpha} - \frac{\partial g}{\partial v^\alpha} 
\frac{\partial f}{\partial w^\alpha}\right)~.
\end{equation}
\subsection{Pseudo-spin coherent states}
\label{ssec:PCS}

We end this section by giving an explicit example of GCS construction, namely those relative to the group $SU(1,1)$\footnote{The Lie group $SU(1,1)$ is defined as 
the group of transformations in the two-dimensional complex plane $\mathbb{C}^2$ that leave invariant the Hermitian form
$\bar{\psi}\psi:=\psi^\dagger\sigma_3\psi=\psi_1^\dagger\psi_1-\psi_2^\dagger\psi_2$, where $\psi=(\psi_1,\psi_2)\in\mathbb{C}^2$ and $\sigma_3$ is the third Pauli
matrix. This group is isomorphic to $SL(2,\mathbb{R})$ and $Sp(2,\mathbb{R})$, and its substantial differences with $SU(2)$ is that it is noncompact and it is not
simply connected. We will study an explicit example of a system related to this group in Sec.~\ref{sec:From_quantum to_classic}.}.
Its generators are the set
$\{\hat{K}_0,\hat{K}_1,\hat{K}_2\}$ which spans the $\mathfrak{su}(1,1)$ algebra
\begin{equation}\label{su(1,1) rules}
[\hat{K}_\alpha,\hat{K}_\beta]=i\epsilon_{\alpha \beta \gamma}\hat{K}^\gamma~,
\end{equation}
where the indices $\alpha,\beta,\gamma \in \{0,1,2\}$ are raised and lowered with the 3-dimensional Minkowski metric $\eta_{\alpha\beta}=\mbox{diag}\{-1,1,1\}$. The
Hilbert space of the system is a unitary irreducible representation of $\mathfrak{su}(1,1)$, which is identified by the so called Bargmann index $k$:
\begin{equation}
\mathcal{H}_k=\{\Ket{k,m},\,m\in\mathbb{N}\;\; k\in\mathbb{R}^+\}~,
\end{equation}
where $\Ket{k,m}$ are the simultaneous eigenstates of $\hat{K}_0$ and of the Casimir operator $\hat{K}^2=-\hat{K}_\alpha\hat{K}^\alpha$ such that
\begin{equation}\label{SU(1,1) eigenstates def}
\begin{cases}
\hat{K}^2\Ket{k,m}=k(k-1)\Ket{k,m}~,\\
\hat{K}_0\Ket{k,m}=(k+m)\Ket{k,m}~.
\end{cases}
\end{equation}
The unitary irreducible representations of $SU(1,1)$ (that are infinite-dimensional since the group is not compact) have been firstly discussed by Bargmann 
\cite{Bargmann47} as incidental to his discussion of the Lorentz group. One can find a consolidated review in Ref. \cite{BiedenharnEA65}. In this paper we will only
refer to the representations of the group $SO(1,2)=SU(1,1)/\mathbb{Z}_2$, obtained by Barut and Fronsdal in Ref. \cite{BarutF65}.\\ To construct GCS we need a
reference state. Given the index-$k$ representation we choose the lowest-weight state i.e. $\Ket{\Phi_0}=\Ket{k,m=0}$ and we identify the stabilizer subgroup
$\mathcal{F}$ by
\begin{equation}\label{SU(1,1)stabilizer}
 e^{i\delta \hat{K}_0}\Ket{k,0}=e^{i\delta k}\Ket{k,0}~~~{\rm with} ~~~\delta \in \mathbb{R}~,
\end{equation}
and hence $\mathcal{F}=U(1)$ .
We can now consider the coset $SU(1,1)/U(1)$ to define the pseudo-spin coherent states (PCS)
\begin{equation}
\Ket{\Omega}=\hat{\Omega}\Ket{k,0}~~\mbox{with}~~\hat{\Omega}\in SU(1,1)/U(1)~,
\end{equation}
where  $\hat{\Omega}$ can be parameterized as:
\begin{equation}\label{SU(1,1) displacement operator}
\hat{\Omega}=\mathrm{e}^{\xi \hat{K}_+ - \xi^* \hat{K}_-}
\end{equation}
with $\hat{K}_\pm=\hat{K}_1 \pm i \hat{K}_2$ shift operators satisfying $[\hat{K}_+,\hat{K}_-]=-2\hat{K}_0$, $[\hat{K}_0,\hat{K}_\pm]=\pm \hat{K}_\pm$.  Points $\Omega$ on the manifold associated to the coset $SU(1,1)/U(1)$ can be identified by the
complex coordinates $(\xi,\xi^*)$; this allows one to express $\hat{\Omega}$, using the standard $(2\times 2)$ matrix representation\footnote{This representation is
finite dimensional and hence not Hermitian.}
\begin{equation}\label{2-dim representation of SU(1,1)}
\hat{K}_+=\begin{bmatrix}
     0 & i\\
                 0 & 0
    \end{bmatrix}\;\;\hat{K}_-=\begin{bmatrix}
                            0 & 0\\
                                        i & 0
                           \end{bmatrix}\;\;\hat{K}_0=\begin{bmatrix}
                                                   \frac{1}{2} &      0      \\
                                                                    0      & -\frac{1}{2}
                                                  \end{bmatrix}~,
\end{equation}
by means of the matrix \cite{ZhangFG90}
\begin{equation}
\begin{bmatrix}
\sqrt{1+\zeta\zeta^*} & \zeta \\
\zeta^* & \sqrt{1+\zeta\zeta^*}
\end{bmatrix}~,
\end{equation}
with $-i\zeta$ defined by Eq. (\ref{zeta coordinates})\footnote{A factor $i$ is needed to define $\zeta$ because the representation (\ref{2-dim representation
of SU(1,1)}) is not Hermitian.}. Introducing ``polar'' coordinates $(\rho,\phi)\in\mathbb{R}\times[0,2\pi]$ via
\begin{equation}
i\xi=\frac{\rho}{2}\mathrm{e}^{-i\phi}
\end{equation}
Eqs. (\ref{zeta coordinates}) and (\ref{tau coordinates}) define the $\zeta$- and $\tau$-coordinates as
\begin{equation}\label{SU(1,1) tau coordinates}
\zeta=\sinh{\frac{\rho}{2}}\mathrm{e}^{-i\phi},\;\;\tau=\tanh{\frac{\rho}{2}}\mathrm{e}^{-i\phi}~.
\end{equation}
Eq. (\ref{SU(1,1) displacement operator}) can be written \cite{Perelomov85} in the $\tau$-coordinates
\begin{equation}
\hat{\Omega}=(1-|\tau|^2)^k\mathrm{e}^{\tau \hat{K}_+}
\end{equation}
so that the natural metric defined in Eq. (\ref{natural metric}) emerges via
\begin{equation}
\begin{aligned}
&\Ket{\tilde{\tau}}=e^{\tau \hat{K}_+}\Ket{k,0}~,~N(\tau,\tau^*)=(1-|\tau|^2)^{-2k}~,\\
&F(\tau,\tau^*)=-2k\log{(1-|\tau|^2)}~,
\end{aligned}
\end{equation}
as
\begin{equation}\label{natural PS2 metric}
ds^2=\frac{2k}{(1-|\tau|^2)^2}d\tau d{\tau}^*=\frac{k}{2}(d^2\rho+\sinh^2{\!\rho}\,d^2\phi)~,
\end{equation}
where $\Ket{\tilde{\tau}}$ is a non-normalized PCS. Moreover, it is possible to show \cite{Perelomov85} that the completeness relation (\ref{coherent state
completeness}) is verified for any $k>1/2$ in the form
\begin{equation}
\begin{aligned}
&\int_{SU(1,1)/U(1)}{d\mu_k(\tau)\,\Ket{\tau}\Bra{\tau}} = \mathbb{I}~,\\
&\mbox{with}~~~d\mu_k(\tau)=\dfrac{2k-1}{\pi}\dfrac{d\tau d\tau^*}{(1-|\tau|^2)^2}~.
\end{aligned}
\end{equation}
The manifold associated to
$SU(1,1)/U(1)$ is called ``Bloch'' pseudosphere $PS^2$ (see for instance Chap.1 of Ref.~\cite{Coppo19MThesis} for further details).

\section{From a quantum theory to a classical one}\label{sec:From_quantum to_classic}
In this section, following Ref.~\cite{Yaffe82}, we show how a large-$N$ limit of a quantum theory can 
formally define a classical dynamics. Let us first specify what makes a theory recognizable as a quantum or a classical one:
as mentioned in Sec.\ref{sec:GCS}, a {\itshape quantum theory} $\mathcal{Q}$ is defined by:
\begin{itemize}
\item a Lie Algebra $\mathfrak{g}$,
\item a Hilbert space $\mathcal{H}$ that carries an irreducible representation of $\mathfrak{g}$,
\item a Hamiltonian operator $\hat{H}\in\mathfrak{g}$.
\end{itemize}
A {\itshape classical theory} $\mathcal{C}$ is instead determined by\footnote{More accurately this is the definition of {\itshape Hamiltonian classical theory}, but
not all classical theories are Hamiltonian. Anyway in this paper we only consider these ones.}:
\begin{itemize}
\item a manifold $\mathcal{M}$,
\item a symplectic form on $\mathcal{M}$, which defines Poisson brackets,
\item a Hamiltonian function $h_{cl}:\mathcal{M}\rightarrow\mathbb{R}$.
\end{itemize}
After the above definitions, one can describe a general procedure for realizing a so-called quantum-to-classical crossover, which is a formal relation between quantum and
classical theories, describing how the first can naturally flow into the latter, possibly when some ``quanticity parameter'' $\chi\in\mathbb{R}^+$ tends to zero.
The limit $\chi\rightarrow 0$ is dubbed ``classical limit'' and, in order to exist, certain conditions must be fulfilled, that isolate the minimal structure that
the starting quantum theory should possess. These conditions are satisfied by a large class of quantum theories, namely the {\itshape Large-N quantum theories} that
feature a global symmetry. If this is the case, $\chi$ is a decreasing function of the number $N$ of degrees of freedom, and $\chi\rightarrow 0$ when
$N\rightarrow \infty$. This reveals that many-variables, provided with a global symmetry, lie behind any quantum-to-classical crossover. \\

\subsection{When does a quantum theory have a classical limit?}\label{ssec:when_quantum_to_classical}
Consider a quantum theory $\mathcal{Q}_\chi$ defined by the Lie algebra $\mathfrak{g}$, the Hilbert space $\mathcal{H}_\chi$ and the Hamiltonian 
$\hat{H}_\chi$. Be such theory characterized by some parameter $\chi$ which is assumed to take positive real values, including the limiting $\chi=0$ one. Once
identified the dynamical group $\mathcal{G}$ of the theory via a Lie exponential map on $\mathfrak{g}$, and its irreducible unitary representation\footnote{Notice
that the abstract group $\mathcal{G}$ and its algebra $\mathfrak{g}$ do not depend on $\chi$, which instead enters $\mathcal{G}_\chi$ and its algebra
$\mathfrak{g}_\chi$ via the $\chi$-dependence of the Hilbert space $\mathcal{H}_\chi$.} $\mathcal{G}_\chi$ on $\mathcal{H}_\chi$, we can construct the GCS
$\Ket{\Omega_\chi}$. They will be in one-to-one correspondence with the points $\Omega_\chi$ of the manifold $\mathcal{M}_\chi$ associated to the coset
$\mathcal{G}/\mathcal{F}_\chi$, where $\mathcal{F}_\chi$ is the stabilizer with respect to a reference state $\Ket{0_\chi}\in\mathcal{H}_\chi$.
For any operator $\hat{A}$ acting on $\mathcal{H}_\chi$ one can define the symbol $A(\Omega_\chi)$ by
\begin{equation}\label{GCS symbol no class}
A(\Omega_\chi):={\Bra{\Omega_\chi}\hat{A}\Ket{\Omega_\chi}}~,~\forall\Omega_\chi\in\mathcal{M}_\chi~.
\end{equation}
As pointed out in Ref.\cite{Yaffe82}, in order to have some control over the limit $\chi\rightarrow 0$, suppose that it is possible to arrange the set of GCS
in the equivalence classes
\begin{equation}
\left[\Ket{\Omega_\chi}\right]_\sim:=\Ket{\Omega}_\chi
\end{equation}
obtained from the equivalence relation
\begin{equation}\label{equivalence relation kappa}
\begin{aligned}
&\Ket{\Omega_\chi}\sim\Ket{\Omega'_\chi}\\
&\mbox{if}\quad \lim_{\chi \to 0} A(\Omega_\chi)=\lim_{\chi \to 0}A(\Omega'_\chi)<\infty~,~
\forall \hat{A}\in\mathcal{K}~,
\end{aligned}
\end{equation}
where, in order to ensure that the limit is well defined, $\mathcal{K}$ is a restricted set of operators satisfying
\begin{equation}\label{classical operator def}
\begin{aligned}
&\lim_{\chi \to 0} \dfrac{\Bra{\Omega}\hat{A}\Ket{\Omega'}_\chi}{\Braket{\Omega|\Omega'}_\chi}:=A(\Omega,\Omega')_\chi<\infty~,\\ 
&\forall \Omega,\Omega'\in\mathcal{M}_\chi/\!\!\sim~;
\end{aligned}
\end{equation}
operators in $\mathcal{K}$ will be called {\it classical} operators. Since the symbols of classical operators upon GCS that belong to a same class are equal, according to Eq. (\ref{equivalence relation kappa}), we will use the notation:
\begin{equation}\label{GCS symbol}
\begin{aligned}
&A_\chi(\Omega):=A(\Omega_\chi)={\Bra{\Omega}\hat{A}\Ket{\Omega}_\chi}~,\\ 
&\forall \Omega\in\mathcal{M}_\chi/\!\!\sim~.
\end{aligned}
\end{equation}
It can be demonstrated \cite{Yaffe82} that, in order for the theory $\mathcal{Q}_\chi$ to have a $\chi\rightarrow 0$ limit that corresponds to a classical theory, 
the following
conditions must hold:
\begin{enumerate}
\item [\textbf{1)}] \textbf{Irreducibility of $\mathcal{G}_\chi$}\\
As mentioned above, each representation $\mathcal{G}_\chi$ of the dynamical group acts irreducibly on the corresponding Hilbert space $\mathcal{H}_\chi$. This
requirement assures that for each $\chi$ the quantum theory is well defined. Using the Schur's lemma and the invariance of the measure on the coset
$\mathcal{G}/\mathcal{F}_\chi$, this assumption implies eq. (\ref{coherent state completeness}), i.e.
\begin{equation}\label{assumption I}
\mathbb{I}_\chi=c_\chi\int_{\left(\rfrac{\mathcal{G}}{\mathcal{F}_\chi}\right)/\sim} {d\mu(\Omega)\,\Ket{\Omega}\Bra{\Omega}_\chi}~,
\end{equation}
where $c_\chi$ is a constant depending on the normalization of the group measure and must be computed explicitly. Notice that the measure $d\mu(\Omega)$ does not 
depend on $\chi$ and hence remains the same as $\chi\rightarrow 0$.
\item [\textbf{2)}] \textbf{Uniqueness of the ``Zero'' operator}\\
The zero operator $\hat{Z}$ is the only one for which $Z_\chi(\Omega)=0\quad\forall\Omega\in \mathcal{M}_\chi/\!\!\sim$. As a consequence, two different 
operators cannot have the same symbol, implying that any operator can be uniquely recovered from its expectation value on GCS, i.e. from its symbols.
\item [\textbf{3)}] \textbf{Exponential decrease of inequivalent coherent states overlap}\\ The overlap between classically inequivalent GCS exponentially
decreases as $\chi\rightarrow 0$, i.e. \begin{equation} \lim_{\chi \to 0} {\Braket{\Omega|\Omega'}_\chi}=\mathrm{e}^{-\lim_{\chi \to
0}{\dfrac{\Delta(\Omega,\Omega')_\chi}{\chi}}} \end{equation} where $\exists\lim_{\chi \to 0}{\Delta(\Omega,\Omega')_\chi}$ $\forall
\Omega,\Omega'\in\mathcal{M}_\chi/\!\!\sim$ and \[
 \operatorname{Re}\Delta(\Omega,\Omega')_\chi  \begin{dcases*}
        > 0  & if $\Ket{\Omega}_\chi\neq\Ket{\Omega'}_\chi$ \\
        = 0  & if $\Ket{\Omega}_\chi=\Ket{\Omega'}_\chi$
        \end{dcases*}
\]
The result is that, when $\chi\rightarrow 0$ inequivalent coherent states become orthogonal, i.e. distinguishable. As a consequence, the factorization
\begin{equation}\label{expectation values limit}
\lim_{\chi \to 0}[(AB)_\chi(\Omega)- A_\chi(\Omega)B_\chi(\Omega)]=0
\end{equation}
holds for any pair $\hat{A}$ and $\hat{B}$ of classical operators.
\item [\textbf{4)}] \textbf{Classical limit of the Hamiltonian}\\
The operator $\chi \hat{H}_\chi$ is a classical operator. This ensures that the coupling constants in the Hamiltonian are scaled in a manner that maintains
sensible dynamics as $\chi\rightarrow 0$, so as to define a meaningful classical limit.
\end{enumerate}
If the hypothesis 1-4 are satisfied there is a phase space on which a classical dynamics can be defined: it is the manifold
$\mathcal{M}=\left(\mathcal{M}_{\chi}/\!\!\sim\right)_{\chi\rightarrow 0}$ whose points $\Omega$ are in one-to-one correspondence with the GCS classes
$\left[\Ket{\Omega_{\chi\rightarrow 0}}\right]_\sim:=\Ket{\Omega}$, and which can be equipped with the natural metric and the symplectic structure defined by Eqs.
(\ref{natural metric})-(\ref{symplectic form}). Using the coordinates $w^\beta,v^\beta$ as in Eq. (\ref{w,v coordinates}) the classical Hamiltonian turns out to be
\cite{Yaffe82}
\begin{equation}\label{definition of classical hamiltonian}
h_{cl}(v^\beta,w^\beta)=\lim_{\chi \to 0} \chi H_\chi (\Omega)~.
\end{equation}

\subsection{Large-$N$ quantum theories: crucial role of global symmetries}\label{Large-N and global symmetries}
All known types of quantum theory $\mathcal{Q}_N$ described by $N$ degrees of freedom (dof) and equipped with a global symmetry $\mathfrak{X}(N)$ are found to satisfy 
the conditions 1-4, where $\chi=\chi(N)$ is a decreasing function such that $\lim_{N\to\infty}\chi(N)=0$. 
The symmetry $\mathfrak{X}(N)$ is called global, for $\mathcal{Q}_N$, if the related transformations act on all its $N$ dof. In fact, the existence of the symmetry $\mathfrak{X}(N)$ is crucial, as it is responsible for the possible reduction of dof defining $\mathcal{Q}_N$, once it has flowed, for $N\rightarrow \infty$, in the corresponding classical theory. Let us hence show how the symmetry plays its role:

* Saying that a theory $\mathcal{Q}_N$ has a certain symmetry implies that all the relevant operators $\hat{A}$ of $\mathcal{Q}_N$ satisfy the relation
$\mathcal{U}\hat{A}\mathcal{U}^\dagger=\hat{A}\quad \forall \mathcal{U}\in \mathfrak{X}(N)~$.

* Considering the GCS of $\mathcal{Q}_N$ and the symbols defined by Eq. (\ref{GCS symbol no class}), it is hence

\noindent
$A(\Omega_N)=\Bra{\Omega_N}\hat{A}\Ket{\Omega_N}=\Bra{\Omega_N}\mathcal{U}\hat{A}\mathcal{U}^\dagger\Ket{\Omega_N}=$
\newline 
$\Bra{\Omega_N^\mathcal{U}}\hat{A}\Ket{\Omega_N^\mathcal{U}}=A(\Omega_N^\mathcal{U})$,
where $\Ket{\Omega_N^\mathcal{U}}:=\mathcal{U}\Ket{\Omega_N}$.

* This suggests to define the equivalence relation:
\begin{equation}
\label{equivrelN}
\Ket{\Omega_N}\sim\Ket{\Omega'_N}~\mbox{if}\quad  A(\Omega_N)=A(\Omega'_N)~,~
\end{equation}
for any relevant operator $\hat{A}$,
in order to arrange the GCS in the classes $\left[\Ket{\Omega_N}\right]_\sim:=\Ket{\Omega}_N$. In this way all the states connected through a symmetry transformation
are equivalent.

* In the limit $N\rightarrow\infty$ only the classical operators defined by Eq. (\ref{classical operator def}) clearly remain relevant and, comparing Eq.
(\ref{equivrelN}) and (\ref{equivalence relation kappa}), one obtains that the points on the classical phase $\mathcal{M}$ are identified by the classes
$\Ket{\Omega}:=\left[\Ket{\Omega_{N\rightarrow\infty}}\right]_\sim$, rather than by the huge number of GCS $\Ket{\Omega_{N\rightarrow\infty}}$.

Finally we remark that \underline{not all} of the Large-$N$ quantum theories flow into classical theories: in order to realize the crossover a global symmetry
is needed. In fact, if such a symmetry is not present the theory will remain quantum also for $N\rightarrow \infty$. For instance consider a theory
describing a Large $\!$- $\!\!N$ set of indistinguishable particles: speaking about some global symmetry is clearly meaningless if one cannot distinguish between a
variable and another one. Indeed it is well known that the quantum effects in a gas of indistinguishable particles are particularly relevant, especially when its
density (for a fixed temperature) is high.
\section{Large $\!$- $\!\!N$ limit of $O(N)$ vector models}\label{O(N) vector model}
Consider a $O(N)$ global invariant quantum theory $\mathcal{Q}_N$ describing a system of $N$ one-dimensional distinguishable spinless particles: its Hamiltonian acts
on the Hilbert space $\mathcal{H}_N$ and can be taken as an arbitrary polynomial\footnote{If $\mathfrak{g}$ is the Lie algebra defining the theory, we consider the
Hamiltonian as an element of the universal enveloping algebra $U(\mathfrak{g})=T(\mathfrak{g})/I$, where
$T(\mathfrak{g})=K\oplus\mathfrak{g}\oplus(\mathfrak{g}\otimes\mathfrak{g})\oplus(\mathfrak{g}\otimes\mathfrak{g}\otimes\mathfrak{g})\oplus\cdots$ is the tensor
algebra of $\mathfrak{g}$ ($K$ is the field over which $\mathfrak{g}$ is defined) and $I$ is the two-sided ideal over $T(\mathfrak {g})$ generated by elements of the
form $\hat{A}\otimes\hat{B}-\hat{B}\otimes\hat{A}-[\hat{A},\hat{B}]$ with $\hat{A},\hat{B}\in\mathfrak{g}$. Informally $U(\mathfrak{g})$ is the algebra of the
polynomials of $\mathfrak{g}$. It is possible to demonstrate \cite{BarutR80} that the representations of $\mathfrak{g}$ and $U(\mathfrak{g})$ are the same.} of the
form
\begin{equation}\label{O(N)-invariant hamiltonian}
\hat{H}_N=N\,h[\hat{A},\hat{B},\hat{C}]~,
\end{equation}
where $\hat{A},\hat{B},\hat{C}$ are the basic $O(N)$ invariants:
\begin{equation}\label{A,B,C basic operators}
\begin{dcases}
\hat{A}  =  \dfrac{1}{2}\sum_{i}\hat{q}_i^2~, \\
\hat{B}  =  \dfrac{1}{2}\sum_{i}(\hat{q}_i\hat{p}_i+\hat{p}_i\hat{q}_i)~,\\
\hat{C}  =  \dfrac{1}{2}\sum_{i}\hat{p}_i^2~,
\end{dcases}
\end{equation}
with positions $\hat{q}_i$ and conjugated momenta $\hat{p}_i$ satisfying the canonical commutation relations:
\begin{equation}\label{commutation relation N particles}
i\,[\hat{p}_i,\hat{q}_j]=\dfrac{1}{N}\delta_{ij}\hat{\mathbb{I}}~,
\end{equation}
with $i,j=1,...,N$ particle index.
Applying the formalism of section \ref{sec:From_quantum to_classic} with $\chi=1/N$, as suggested in Ref.~\cite{Yaffe82}, one finds the classical limit of the $\mathcal{Q}_N$ for $N\rightarrow \infty$ as follows\footnote{
In order to avoid
explicit rescalings of the coupling constants in the Hamiltonian as $N\rightarrow\infty$, a factor $1/\sqrt{N}$ has been included in the definition of $\hat{q}_i$
and $\hat{p}_i$, as seen from Eq. (\ref{commutation relation N particles})}.

\textbf{Identification of the dynamical group and coherent states}\\
The dynamical group $\mathcal{G}_N$ is defined as the group generated by the operators $\hat{A}$, $\hat{B}$ and $\hat{C}$.
From (\ref{commutation relation N particles}) we obtain the commutation rules for its Lie algebra $\mathfrak{g}_N$
\begin{equation}\label{A,B,C commutation rules}
[\hat{A},\hat{B}]=\dfrac{2i}{N}\hat{A}~,~[\hat{A},\hat{C}]=\dfrac{i}{N}\hat{B}~,~[\hat{B},\hat{C}]=\dfrac{2i}{N}\hat{C}~,
\end{equation}
that, via the linear transformations
\begin{equation}\label{K_alpha transformations}
\hat{K}_0=\frac{1}{2}(\hat{A}+\hat{C})~,~\hat{K}_1=\frac{1}{2}\hat{B}~,~\hat{K}_2=\frac{1}{2}(\hat{A}-\hat{C})~,
\end{equation}
become Eqs. (\ref{su(1,1) rules}), with the structure constants consistently rescaled by a factor $1/N$ (in fact the unscaled rules are realized with the elements
$N\hat{K}_\alpha$). $\mathcal{G}_N$ can then be regarded as a unitary representation on $\mathcal{H}_N$ of the group $\mathcal{G}=SU(1,1)$ and the GCS for
$\mathcal{Q}_N$ are the PCS introduced in Sec. \ref{ssec:PCS}: For convenience, the indices $k$ (Bargmann index) and $m$, there defined will be rescaled by
a factor $N$, i.e. $k\rightarrow Nk$, $m\rightarrow Nm$.

\textbf{Classical operators and equivalent states}\\
Using the $\tau$-coordinates introduced at the end of section \ref{ssec:PCS}, the PCS overlaps \cite{Novaes04}
\begin{equation}\label{PCS overlaps}
\Braket{\Omega | \Omega'}_N=\dfrac{(1-|\tau'|^2)^{Nk} (1-|\tau|^2)^{Nk}}{(1-\tau'\tau^*)^{2Nk}}~,
\end{equation}
and the matrix elements
\begin{equation}\label{K0,K1,K2 matrix element}
\begin{dcases}
\dfrac{\Bra{\Omega}\hat{K_0}\Ket{\Omega'}}{\Braket{\Omega | \Omega'}}=k\dfrac{1+\tau'\tau^*}{1-\tau'\tau^*}~, \\
\dfrac{\Bra{\Omega}\hat{K_1}\Ket{\Omega'}}{\Braket{\Omega | \Omega'}}=2k\dfrac{\operatorname{Re}\tau}{1-\tau'\tau^*}~,\\
\dfrac{\Bra{\Omega}\hat{K_2}\Ket{\Omega'}}{\Braket{\Omega | \Omega'}}=-2k\dfrac{\operatorname{Im}\tau}{1-\tau'\tau^*}~,
\end{dcases}
\end{equation}
show us that $\hat{K_\alpha}$ are all classical operators, in agreement with the definition (\ref{classical operator def}); therefore, as they are basic operators,
all $O(N)$-invariant operators are classical.
We can obtain Eqs. (\ref{K0,K1,K2 matrix element}) using the action of $\hat{K}_\pm$ on the $\Ket{k,m}$-states, defined by Eq.
(\ref{SU(1,1) eigenstates def}),
\begin{equation*}
\begin{aligned}
& \hat{K}_\pm \Ket{k,m} = \\
&=\sqrt{N(k+m)(N(k+m)\pm 1)-Nk(Nk-1)}\Ket{k,m\pm 1}~,
\end{aligned}
\end{equation*}
and the PCS expansion in terms of $\Ket{k,m}$, which in $\tau$-coordinates is \cite{Novaes04}
\begin{equation}
\begin{aligned}
\Ket{\Omega}=&(1-|\tau|^2)^{Nk}\cdot \\
&\cdot \sum_{m=0}^\infty{\sqrt{\frac{\Gamma(N(2k+m))}{(Nm)!\Gamma(2Nk)}}}\tau^{Nm} \Ket{k,m}~,
\end{aligned}
\end{equation}
where $\Gamma$ is the Euler's gamma function.
If we now consider Eq. (\ref{K0,K1,K2 matrix element}) for $\tau'=\tau$, i.e.
\begin{equation}\label{K0,K1,K2 expectation values}
\begin{dcases}
K_0(\Omega)  =  \Bra{\Omega}\hat{K_0}\Ket{\Omega}  =  k\dfrac{1+|\tau|^2}{1-|\tau|^2} \\
K_1(\Omega)  =  \Bra{\Omega}\hat{K_1}\Ket{\Omega}  =  2k\dfrac{\operatorname{Re}\tau}{1-|\tau|^2}\\
K_2(\Omega)  =  \Bra{\Omega}\hat{K_2}\Ket{\Omega}  =  -2k\dfrac{\operatorname{Im}\tau}{1-|\tau|^2}
\end{dcases}~,
\end{equation}
we correctly find that the symbols of classical operators are different only for states belonging to different equivalence classes.

\textbf{Proof of hypothesis}\newline
 {\itshape 1) Irreducibility of $\mathcal{G}_N$}: 
This hypothesis needs no proof, as we actually assume it in order to define a consistent quantum theory for any fixed $N$. Notice, in fact, that we have
already enforced it when considering the PCS, as we have required the value of the Casimir operator 
\noindent $\hat{K}^2=-\hat{K}_\alpha \hat{K}^\alpha$ on $\mathcal{H}_N$ to
be fixed (Schur's lemma).\newline
{\itshape 2) Uniqueness of the ``Zero" operator}:
Suppose there exists some operator $\hat{Z}$ for which $Z(\Omega)=\Bra{\Omega}\hat{Z}\Ket{\Omega}=0$ for any PCS $\Ket{\Omega}$. Using the commutation relations
$[\hat{K}_-,\hat{K}_+]=(2/N)\hat{K}_0$ and $\hat{K}_-\Ket{k,0}=0$, where $\hat{K}_\pm$ and $\Ket{k,0}$ are the shift operators and the reference state, respectively,
introduced in Sec. \ref{ssec:PCS}, one can show by an induction argument\footnote{Since $\Ket{k,0}$, as reference state (see Sec. \ref{ssec:PCS}), is a PCS, $Z(\Omega)=0$
implies $\Bra{k,0}\hat{Z}\Ket{k,0}=0$. Then, assuming that
\begin{equation}\label{K polynomial}
\Bra{k,0}\hat{K}_-...\hat{K}_-\hat{Z}\hat{K}_+...\hat{K}_+\Ket{k,0}=0
\end{equation}
is true when the total number of $\hat{K}_-$ plus $\hat{K}_+$ is less that $n$, we must prove the same holds when such number becomes $n$. Firstly note that
$Z(\Omega)=0$ implies $\Bra{k,0}[[\hat{Z},\hat{\Lambda}_1],\hat{\Lambda}_2],...\hat{\Lambda}_n]\Ket{k,0}=0\;\;\hat{\Lambda}_i\in {\hat{K}_-,\hat{K}_+}$ (choose
$\Ket{\Omega}=\hat{\Omega}\Ket{0}$ with $\hat{\Omega}=\mathrm{e}^{t_1\hat{\Lambda}_1}\mathrm{e}^{t_2\hat{\Lambda}_2}...\mathrm{e}^{t_n\hat{\Lambda}_n}$ and
differentiate $Z(\Omega)$ with respect to each $t_i$). Expanding the multiple commutator, we find that only one term contains all $\hat{K}_-$ operators to the left
and all $\hat{K}_+$ operators to the right of $\hat{Z}$. Every other term contains at least one $\hat{K}_-$ operator which may be pushed right until it annihilates
$\Ket{k,0}$, or one $\hat{K}_+$ operator which may be pushed left. This process produces also commutator terms $[\hat{K}_-,\hat{K}_+]$ which reduce $n$ by two. In
the end the vacuum expectation value of a multiple commutator contains a term of the form (\ref{K polynomial}) with $n$ operators plus lower-order terms which vanish
for induction. Therefore Eq. (\ref{K polynomial}) also holds for a number $n$ of $\hat{K}_-$ plus $\hat{K}_+$ operators.}, that \\
$\Bra{k,0}\hat{K}_-...\hat{K}_-\hat{Z}\hat{K}_+...\hat{K}_+\Ket{k,0}=0$ for any number of $\hat{K}_-$ or $\hat{K}_+$. Polynomials in $\hat{K}_+$ applied to
$\Ket{k,0}$ clearly form a dense set of $O(N)$-invariant states, then $\hat{Z}$ must be the zero operator.\newline
{\itshape 3) Exponential decrease of inequivalent coherent states overlap}:
This condition easily follows from Eq. (\ref{PCS overlaps}), implying
\begin{equation}
\lim_{N \to \infty} {\Braket{\Omega|\Omega'}_N}=\mathrm{e}^{-\lim_{N \to \infty}{N\Delta(\tau,\tau')_N}}~,
\end{equation}
where
\begin{equation}
\begin{aligned}
&\Delta(\tau,\tau')_N=\Delta(\tau,\tau')=\\
&-k\left[\ln{(1-|\tau'|^2)}+\ln{(1-|\tau|^2)}-2\ln{(1-\tau'\tau^*)})\right]~,
\end{aligned}
\end{equation}
so that $\exists\lim_{N \to \infty}{\Delta(\tau,\tau')_N}\;\, \forall \tau,\tau'$ with
\[
 \operatorname{Re}\Delta(\tau,\tau')  \begin{dcases*}
        > 0  & if $\Ket{\Omega}\neq\Ket{\Omega'}$ \\
        = 0  & if $\Ket{\Omega}=\Ket{\Omega'}$
        \end{dcases*}
\]\newline
{\itshape 4) Classical limit of Hamiltonian}:
As any $N$-inde\-pe\-ndent polynomial in $\hat{A}$, $\hat{B}$ and $\hat{C}$ is a classical operator, this holds true also for any Hamiltonian of the form
(\ref{O(N)-invariant hamiltonian}).

\textbf{Classical theory}\\
We can define a classical dynamics on the coset $SU(1,1)/U(1)$, which is the manifold $PS^2$ described in 
Sec.\ref{ssec:PCS}, that can be mapped to the so-called Poincar\'e half plane $\mathbb{H}$ by the conformal
transformation
\begin{equation}\label{conformal transformation}
z:=\varrho-iv:=\dfrac{i+\tau}{i-\tau}~.
\end{equation}
$\mathbb{H}$ is endowed with the natural metric 
\begin{equation}\label{EADS2 metric}
ds^2=\dfrac{R^2}{\varrho^2}(d\varrho^2+dv^2)~,
\end{equation}
where $R=\sqrt{k/2}$ , and with the standard Poisson brackets
\begin{equation}\label{EADS2 PB}
\left\{f,g\right\}_{PB}=\frac{\partial f}{\partial v} \frac{\partial g}{\partial w} - \frac{\partial g}{\partial v} \frac{\partial f}{\partial w}~,
\end{equation}
once $w=k/\varrho$ has been defined.
Considering the
transformations (\ref{K_alpha transformations}) together with Eq. (\ref{K0,K1,K2 expectation values}) in the coordinates $(v,w)$, as defined above, we obtain
\begin{equation}\label{A,B,C expectation values w,v variables}
\begin{dcases}
A(v,w)  =   w~, \\
B(v,w)  =  2vw~,\\
C(v,w)  =  w\left(\dfrac{k^2}{w^2}+v^2\right)~.
\end{dcases}
\end{equation}
From Eqs. (\ref{definition of classical hamiltonian}) and (\ref{O(N)-invariant hamiltonian}) we then get the classical Hamiltonian:
\begin{equation}\label{classical O(N) hamiltonian}
h_{cl}(v,w)=h\left(w,2vw,v^2w+k^2 w^{-1}\right)~.
\end{equation}
Finally we can define a classical action
\begin{equation}\label{classical O(N) action}
S_{cl}=\int dt\,\left[v\dot{w}-h_{cl}(v,w)\right]~,
\end{equation}
from which the equations of motion of the classical theory, i.e the Hamilton's equations, can be derived
\begin{equation}
\begin{dcases}
\dot{v}=-\dfrac{\partial h_{cl}}{\partial w}=\left\{h_{cl},v\right\}_{PB}~,\\
\dot{w}=\dfrac{\partial h_{cl}}{\partial v}=\left\{h_{cl},w\right\}_{PB}~.
\end{dcases}
\end{equation}
\textbf{Role of the symmetry $O(N)$}\\
We highlight that in the above construction the role of the global symmetry $O(N)$ is crucial. Indeed had it been absent, the GCS would not have been in one-to-one
correspondence with the points of $PS^2$, but with those of the much bigger $N$-dimensional complex plane $\mathbb{C}^N$. Denoting with $H_4$ the so called Heisenberg group, from which one obtains the standard "Harmonic-oscillator" coherent states \cite{ZhangFG90}, we can graphically summarize the job done
by the symmetry as follows:
\begin{align*}
  &\Ket{\Omega_N}\;\;\;\;\leftrightarrow\;\;\;\; \left(H_4/U^2(1)\right)^N\simeq \mathbb{C}^N,\; \Omega_N\in\mathbb{C}^N \\
    &\;\;\;\Bigg\Downarrow \mbox{$O(N)$-symmetry} \\
  &\Ket{\Omega}_N{:=}\left[\Ket{\Omega_N}\right]_\sim \leftrightarrow {SU(1,1)\over U(1)}\simeq PS^2,\; \Omega \in PS^2
\end{align*}
where the equivalence relation is constructed thanks to the symmetry transformations, as in Eq. (\ref{equivrelN}).

\subsection{Quantum-to-classical crossover of angular momentum}\label{quantum-to-classic angular momentum}
Thanks to Eqs. (\ref{expectation values limit}) and (\ref{K0,K1,K2 expectation values}) we can calculate the expectation value on PCS of the Casimir operator
$\hat{K}^2=-\hat{K}_\alpha\hat{K}^\alpha$:
\begin{equation}\label{K^2 expectation value}
K^2(v,w)=k^2~.
\end{equation}
Notice that this is constant, i.e. it does not depend on the conjugated variables $(v,w)$. This means that it has to flow to a conserved quantity in the classical
motion. It is then a due question to ask: ``which one?''. It is suggestive to analyze the connection of $\hat{K}^2$ with the $N$ degrees of freedom. After some
calculations we obtain:
\begin{equation}\label{Casimir as N-dim angular momentum}
\hat{K}^2=\dfrac{1}{4}\left(\hat{L}^2+\dfrac{1}{4}-\dfrac{1}{N}\right)~,
\end{equation}
where
\begin{equation}
\begin{dcases}
\hat{L}^2=\frac{1}{2}\sum_{ij}\hat{L}_{ij}^2\;\;\;\;\;i,j=1,...,N~,\\
\hat{L}_{ij}=\hat{q}_i\hat{p}_j-\hat{q}_j\hat{p}_i~,
\end{dcases}
\end{equation}
are the modulus and the components of the angular momentum, respectively, of the $N$ degrees of freedom. Then the above mentioned conserved classical quantity might
be an angular momentum. Such an identification is reinforced by the following argument:
\begin{itemize}
\item {\it i)} If we express the positions $\hat{q}_i$ and the momenta $\hat{p}_i$ in terms of the ladder operators $(\hat{a}_i,\hat{a}^\dagger_i)$:
\begin{equation}
\begin{dcases}
\hat{q}_i=\frac{1}{\sqrt{2}}(\hat{a}^\dagger_i+\hat{a}_i)\\
\hat{p}_i=\frac{i}{\sqrt{2}}(\hat{a}^\dagger_i-\hat{a}_i)
\end{dcases}
\end{equation}
the canonical commutation rules (\ref{commutation relation N particles}) become
\begin{equation}\label{ladder commutation relation N particles}
[\hat{a}_i,\hat{a}_j^\dagger]=\dfrac{1}{N}\delta_{ij}\hat{\bbone} \mbox{ with } i,j=1,...,N
\end{equation}
and thanks to Eqs. (\ref{A,B,C basic operators}) and (\ref{K_alpha transformations}) it is:
\begin{equation}\label{relation K_0 and N}
\hat{K}_0=\dfrac{1}{2}\left(\hat{\mathcal{N}}+\dfrac{1}{2}\right)
\end{equation}
where $\hat{\mathcal{N}}=\sum_{i}\hat{a}^\dagger_i\hat{a}_i$ is the number operator.

\item {\it ii)} As it is well known, the relations (\ref{ladder commutation relation N particles}) imply that the spectrum of $\hat{\mathcal{N}}$ is the set $\mathbb{N}$ of
the natural numbers and then, according to Eq. (\ref{relation K_0 and N}), the spectrum of $\hat{K}_0$ is $\{\frac{1}{2}\left(n+\frac{1}{2}\right),\allowbreak\;\,
n\in\mathbb{N}\}$. Defining $n:=l+2m$ with $l,m \in \mathbb{N}$, and considering Eqs. (\ref{SU(1,1) eigenstates def}), the possible values of $k$ must be
\begin{equation}\label{k-l relation}
k=\dfrac{\tilde{l}}{2}:=\dfrac{1}{2}\left(l+\dfrac{1}{2}\right)\;\;\mbox{ with } l\in\mathbb{N}~;
\end{equation}
in fact, a more exhaustive demonstration of Eq. (\ref{k-l relation}) can be found in Ref. \cite{Friedmann12}.
\item {\it iii)} The physical meaning of the natural number $l$ is revealed when inserting Eq. (\ref{k-l relation}) in Eq. (\ref{SU(1,1) eigenstates def}) to see that, using
Eq. (\ref{Casimir as N-dim angular momentum}), it is\footnote{Notice that, considering the rescaling of the $SU(1,1)$ commutation rules and of the indices $k$, $m$,
Eq. (\ref{SU(1,1) eigenstates def}) assumes the form:
\begin{equation*}\label{SU(1,1) eigenstates def rescaled}
\begin{cases}
N^2\hat{K}^2\Ket{k,m}=Nk(Nk-1)\Ket{k,m}\\
N\hat{K}_0\Ket{k,m}=N(k+m)\Ket{k,m}
\end{cases}~
\end{equation*}}
\begin{equation}\label{spectrum of 3d angular momentum}
\begin{aligned}
&\hat{L}^2\Ket{l,m}=l\left(l+1-\dfrac{2}{N}\right)\Ket{l,m}\\
&\xhookrightarrow[N\rightarrow \infty]{}\\
&\hat{L}^2\Ket{l,m}=l(l+1)\Ket{l,m}\;\;\; l\in\mathbb{N}
\end{aligned}
\end{equation}
where we have written the eigenstates $\Ket{k,m}$ as $\Ket{l,m}$. Eq. (\ref{spectrum of 3d angular momentum}) shows that, if the limit $N\rightarrow \infty$ is
performed, the operator $\hat{L}^2=4\hat{K}^2-\frac{1}{4}$ has the same spectrum of the modulus of a 3-dimensional orbital angular momentum operator.
\item {\it iv)} Finally inserting Eq. (\ref{k-l relation}) in Eq. (\ref{K^2 expectation value}) it is
\begin{equation}\label{Casimir as angular momentum}
4K^2(v,w)=4k^2=\tilde{l}^2~;
\end{equation}
it is hence appropriate to assume that the quantity $4k^2$ flows to a classical 3d angular momentum, conserved in the motion. Moreover the dependence on
$\tilde{l}=l+\frac{1}{2}$ confirms that the limit $N\rightarrow \infty$ is a classical one.\footnote{The classical limit of a spin-$j$ system can be naively
implemented substituting the spin operators with classical vectors that freely move on a sphere of radius $(j+1/2)$ \cite{Lieb73}.}
\end{itemize}
In the end, a ``bridge'' from ``quantum to classical'' is built thanks to the three real parameters $(v,w,k)$ that have a genuine quantum origin but, in the limit
$N\rightarrow \infty$, entering in the Hamiltonian (\ref{classical O(N) hamiltonian}), do acquire a proper classical nature. In particular, from the quantum
viewpoint, the Bargmann index $k$ identifies the theory's Hilbert space $\mathcal{H}_N$ as an irreducible representation of $SU(1,1)$ and $(v,w)$ the overcomplete
set of PCS. As a result of the crossover to the classical theory, $(v,w)$ become the conjugated variables defining the motion, and $4k^2$ is a conserved angular
momentum.\\

$\!\!\!\!\!\!\!\!\!\!$Despite the above, quite convincing, discussion, there is a caveat: the action (\ref{classical O(N) action}) tells us that the classical motion
is 1-dimensional, implying that no angular momentum can be defined. The only possibility is hence that such an angular momentum is external to the system. In fact,
noticing that $w$ has to be positive \footnote{This is easily proved by noticing from Eq. (\ref{SU(1,1) tau coordinates}) that $|\tau|^2<1$, and using the transformation Eq.(\ref{conformal transformation}) and the one after Eq. (\ref{EADS2 PB})}, it can be mapped into a radius $r$; therefore, we suggest that
the emerging classical theory describes the central motion of a 3-dimensional particle in the 1-dimensional effective-potential formalism, with respect to the radial coordinate $r$. In
the next subsection we show how this statement is substantiated.
\subsection{A paradigmatic example: the free particles}\label{Free particles}

Let us use the above designed procedure to find the classical limit of a quantum theory 
that describes a number $N\rightarrow\infty$ of one-di\-men\-sional
distinguishable free particles. The quantum Hamiltonian is:
\begin{equation}\label{Free quantum Hamiltonian}
\hat{H}_N=\frac{N}{2}\sum_{i} \hat{p}_i^2=N\,\hat{C}~.
\end{equation}
The classical phase space is $\mathbb{H}$ with the two coordinates $(v,w)$; the classical Hamiltonian describing the limit $N\rightarrow\infty$ is, according to Eq.
(\ref{classical O(N) hamiltonian}),
\begin{equation}\label{Classical limit of free Hamiltonian in v,w}
h_{cl}(v,w)= v^2w + k^2 w^{-1}~.
\end{equation}
If we relabel $w=r^2/2,\;v=p/r$, then $\left\{p,r\right\}_{PB}=1$ and Eq. (\ref{Classical limit of free Hamiltonian in v,w}) becomes
\begin{equation}\label{Classical limit of free Hamiltonian in p,r}
h_{cl}(p,r)=\dfrac{p^2}{2}+\dfrac{\tilde{l}^2}{2r^2}~.
\end{equation}
This is indeed the Hamiltonian of a classical 3-d free particle with angular momentum $L^2=\tilde{l}^2$ in the effective potential formalism.\\
We have thus obtained that a large number $N$ of quantum free particles corresponds to one single classical rotating particle. It is of great relevance and
significance that one cannot recover the quantum Hamiltonian (\ref{Free quantum Hamiltonian}) from the classical one (\ref{Classical limit of free Hamiltonian in
p,r}) simply substituting the dynamical variables $p^2$ and $r^2$ with the operators $\hat{p}^2=\sum_i{\hat{p_i}^2}$ and $\hat{r}^2=\sum_i{\hat{q_i}^2}$, and
imposing the rules $i[\hat{p},\hat{r}]=\hat{\mathbb{I}}$ i.e. by a naive ``quantization'': indeed the classical limit of quantum theory is most often a completely
different theory.

\section{Conclusions}\label{Conclusions}

Before presenting our concluding remarks, let us briefly comment upon the difference between the quantum to classical crossover of a theory, which is the topic to which this work is dedicated, and the
suppression of quantum features in the behaviour of non-isolated quantum systems, either closed or open\footnote{Despite the ambiguity of the terminology, in recent literature "closed" systems are defined as quantum non-isolated systems with an  environment that enters the analysis as a classical-like agent, such as an external magnetic field, or a classical thermal bath. "Open" quantum systems, instead, are those whose environment is, and must be treated as, a quantum system, which implies having to consider phenomena such as entanglement generation, back-flow of information, non-markovianity, and many others.}.
The two processes are of a profoundly different nature.
The former occurs whenever the system that the theory describes is "big" (i.e. made of a very large number of components) and features some global symmetry, irrespective of the possible presence of other systems in the overall setting. It is a process that changes the theoretical framework into which the description of the system is set.
The latter, instead, is observed in quantum systems WITH an environment, and the loss of quantum features is a direct consequence of such an environment being "big" in the above sense. In this case, the principal system stays "quantum" (i.e. described by the formal tools of quantum mechanics), and it can possibly recover its quantum features.
In fact, it has been recently shown\cite{BrandaoPH15,2019FHMV} that the loss of quantum features in systems that interact with environments made by a very large number of components can be formally derived and related with the theory of quantum measurements, once the environment undergoes a quantum to classical crossover as described by the first process.
In brief, one can say that the quantum to classical crossover is defined as a process "per s\'e",
while the loss of quantum features in a system is a consequence of one such crossover occuring in the environment.

We finally get to our Conclusions, and start  by interpreting the results of the previous section from 
a different viewpoint. Consider a quantum theory $\mathcal{Q}_\chi$ that describes a spinless
particle in 1-dimension, and depends on some ``quanticity'' parameter $\chi$, say a coupling constant. Its dynamical group is $H_4$, that can be identified with
$\mathcal{G}_{N=1}$.\footnote{This result is clear when considering Eqs. (\ref{ladder commutation relation N particles}) for $N$=1.} Noting that $H_4\simeq
\mathcal{G}_N$ (they are both representations of $\mathcal{G}$), we find that there exists a correspondence between the GCS $\Ket{\Omega}_\chi$ of $Q_\chi$ and
the GCS $\Ket{\Omega_N}$ of $\mathcal{Q}_N$, i.e. one can write $\Omega=\Omega(\Omega_N)$. Then, given the quantum Hamiltonian $\hat{H}_\chi$ of
$\mathcal{Q}_\chi$, we can find a Hamiltonian $h$ of the form (\ref{O(N)-invariant hamiltonian}) such that
\begin{equation}
H_\chi(\Omega(\Omega_N))=N\,h[A(\Omega_N),B(\Omega_N),C(\Omega_N)]~.
\end{equation}
Therefore, when $N\rightarrow\infty$ we obtain a classical theory not only for $\mathcal{Q}_N$ 
but also for $\mathcal{Q}_\chi$, possibly when $\chi\rightarrow
0$. This argument can be generalized to the $N\rightarrow \infty$ limit of essentially all physical 
quantum theory equipped with a global symmetry.\footnote{Yaffe demonstrates that not only $O(N)$ vector models have a classical limit when $N\rightarrow \infty$, but also $U(N)$ matrix models and $U(N)$-lattice gauge theories.}

One may thus presume that any system we perceive as classical is in fact a particular macroscopic quantum system, of which we observe the effective behaviour. The explicit implementation of Yaffe's procedure given above for a few 
paradigmatic physical systems, clearly illustrate some inherent properties of the quantum to classical crossover in the macroscopic limit, which are worth to be highlighted. 
Among them, probably one of the most apparent is that many different quantum theories can flow into the same classical theory, whose Hamiltonian may appear rather different from the one expected by doing a na\"\i ve classical 
limit of the quantum one. However, the one outcome upon which we would like to comment the most, is that the classical theory we finally get by the present formal approach is not that whose conventional quantization would lead to the quantum theory from which we started: This is, in our opinion, an especially relevant observation, as it tells us that apparently
the proper way of reasoning is that of moving from the quantum to the classical description, and not that of quantizing classical theories. We think that such change of perspective could be fruitful to address some still unsolved problems, as, e.g., a proper quantum description of gravitation.

\begin{acknowledgements}
Financial support from the Universit\`a degli Studi di Firenze in the 
framework of the University Strategic Project Program 2015 (project
BRS00215) is gratefully acknowledged. PV has worked in the framework of the Convenzione Operativa between the Institute for Complex Systems of
the Italian National Research Council (CNR) and the Physics
and Astronomy Department of the University of Florence.
\end{acknowledgements}

\section*{Conflict of interest}

The authors declare that they have no conflict of interest.



%
%

\end{document}